\documentclass[preprint,journal]{vgtc}            

\onlineid{0}

\vgtccategory{Research}

\vgtcpapertype{design study}

\title{Visualizing Historical Book Trade Data: An Iterative Design Study with Close Collaboration with Domain Experts}

\author{%
  \authororcid{Yiwen Xing}{0000-0003-1521-6616},
  \authororcid{Cristina Dondi}{0000-0001-9478-216X},
  \authororcid{Rita Borgo}{0000-0003-2875-6793},
  and
  \authororcid{Alfie Abdul-Rahman}{0000-0002-6257-876X}
}

\authorfooter{
   \item
  	Yiwen Xing, Rita Borgo, and Alfie Abdul-Rahman are with King's College London.
  	E-mail: [yiwen.xing, rita.borgo, alfie.abdulrahman]@kcl.ac.uk.
   \item
  	Cristina Dondi is with University of Oxford. 
        E-mail: cristina.dondi@mod-langs.ox.ac.uk. 
}

\abstract{%
    The circulation of historical books has always been an area of interest for historians. However, the data used to represent the journey of a book across different places and times can be difficult for domain experts to digest due to buried geographical and chronological features within text-based presentations. This situation provides an opportunity for collaboration between visualization researchers and historians.
    This paper describes a design study where a variant of the Nine-Stage Framework~\cite{sedlmair2012design} was employed to develop a Visual Analytics (VA) tool called \emph{DanteExploreVis}. This tool was designed to aid domain experts in exploring, explaining, and presenting book trade data from multiple perspectives. We discuss the design choices made and how each panel in the interface meets the domain requirements. We also present the results of a qualitative evaluation conducted with domain experts.
    The main contributions of this paper include: 1) the development of a VA tool to support domain experts in exploring, explaining, and presenting book trade data; 2) a comprehensive documentation of the iterative design, development, and evaluation process following the variant Nine-Stage Framework; 3) a summary of the insights gained and lessons learned from this design study in the context of the humanities field; and 4) reflections on how our approach could be applied in a more generalizable way.

}


\teaser{
  \centering
  \includegraphics[width=\linewidth]{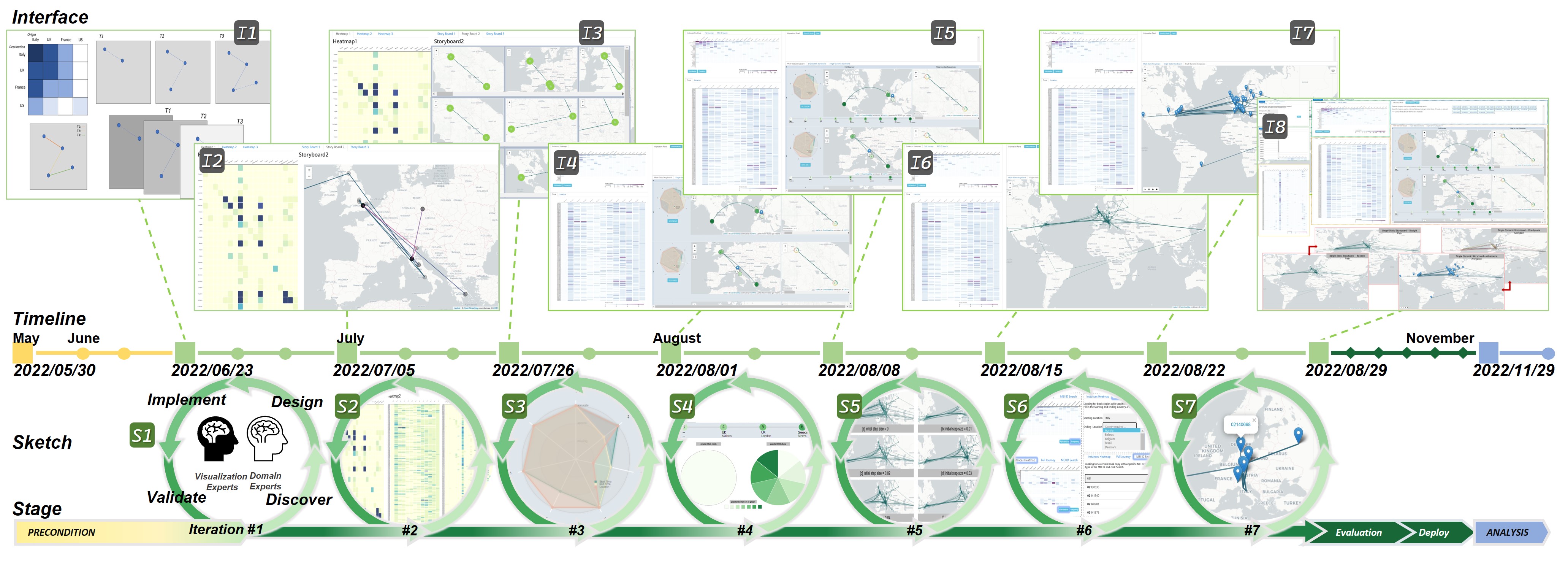}
  \caption{Timeline of the iterative design process, showcasing the evolution of interface changes (I1 -- I8), notable sketches (S1 -- S7) from each iteration, and corresponding stages in the framework variant. This figure provides a visual summary of the design progression and highlights the milestones in the development of \emph{DanteExploreVis}.}
  \label{fig:teaser}
}

\graphicspath{{figs/}{figures/}{pictures/}{images/}{./}} 

\usepackage{epstopdf}
\usepackage{dirtytalk}

\newcommand{\rev}[1]{\textcolor{black}{#1}}

\begin{document}


\firstsection{Introduction}

\maketitle

The trade and circulation of historical books have long been a matter of interest to historians. A book printed in the 15$^{th}$-century has bounced around for centuries, moving from country to country, bearing witness to countless histories of the time, and eventually taking on its present form. By examining book trade data, historians can gain insights into the life and trading history of books, encompassing all copies of a specific printed edition of a literary work. This information can offer historians valuable evidence to interpret various historical phenomena and provide fresh perspectives on pressing issues within the discipline.
In the big data era, tracking the circulation of books is more manageable, but not for historical books. Historians have spent considerable time collecting and integrating historical book records. By examining features such as manuscript annotations, decorations, and binding styles, historians can trace the spatial and temporal movement of these books. The MEI (Material Evidence in Incunabula) database~\cite{mei_2015} compiles these fragmented records, offering valuable data for historical book researchers. However, the absence of appropriate visualization tools creates difficulties in analyzing and interpreting the large dataset, prompting our design study and collaboration with historians.

Prompted by the demand for visualizing book trade data, we initiated an interdisciplinary collaboration with historians in 2021. The \emph{BookTracker} platform was established to develop visualization tools to address various domain needs.
Our previous design study is detailed in \cite{xing2022design}, while this paper focuses on the development of \emph{DanteExploreVis}. Building on our previous experience, we adapt the core phase of the Nine-Stage Framework~\cite{sedlmair2012design} to better suit the iterative nature and prioritize the continuous refinement of domain problems and tasks.
With the adapted Nine-Stage Framework, we crafted a tool capable of fulfilling domain tasks. Throughout the design process, we maintained frequent communication with domain experts, promptly receiving feedback and evaluations after each implementation. The qualitative evaluations of \emph{DanteExploreVis} provided positive feedback.

In this paper, we present the result of the design study conducted with historical book researchers following a variant of the Nine-Stage Framework. During the iterative process, prototypes were refined and enhanced to better address domain tasks, ultimately leading to the development of \emph{DanteExploreVis}. Our contributions include:
\begin{itemize}[noitemsep,nolistsep,leftmargin=*]
    \item Completing a six-month iterative design process with the leading domain expert, resulting in the development of \emph{DanteExploreVis}.
    \item Evaluating the usability and usefulness of \emph{DanteExploreVis} through surveys and expert interviews with a group of domain experts.
    \item \rev{Adapting the Nine-Stage Framework with a more explicit prominence of its iterative essence.}  
    \item Documenting and visualizing the design study process, sharing insights and lessons learned from collaboration with humanities scholars, and discussing the scalability of our approach to wider domains.
\end{itemize}



\section{Domain Background and Data}
\label{sec:Sec2}

%

\subsection{15cBOOKTRADE Project and MEI Database}
The development of \emph{DanteExploreVis} is motivated by the growing volume of records in the MEI database~\cite{mei_2015} and the unmet research needs of the 15cBookTrade Project~\cite{15cBOOKTRADE:2021:web}.

\textbf{15cBOOKTRADE Project} aims to investigate the impact of the European printing revolution (1450-1500) on the development of European civilization during the early modern period. The project utilizes tangible evidence from thousands of surviving 15$^{th}$-century printed books to address fundamental questions about the dissemination of printing in the West, which had previously remained unanswered due to a lack of evidence and inadequate tools for exploring existing data.
The project uses material, documentary, and bibliographical evidence to reconstruct the history of each book, including its provenance or `life' from the time of its printing to its current location~\cite{dondi201315cbooktrade}. Each piece of evidence is recorded in a separate block of provenance, which is tagged geographically and chronologically. The `life' of a book is therefore represented in MEI records as a sequence of provenance blocks.
Mobility is an inherent characteristic of printed books, and the success of the book trade relied on the distribution of hundreds of copies of each edition beyond the place of production. Therefore, studying the book trade is essential to understanding the printing revolution. 

\textbf{MEI}~\cite{mei_2015} has been developed to provide a physical representation of the circulation of books over time, from their production location to their current locations: the possibility to visualize the movement of books has always been a priority, firstly to advance scholarship by analyzing the extensive data and identifying trends, and secondly, to effectively communicate research findings to the general public.
The MEI database is a critical output of the 15cBOOKTRADE Project, which consolidates evidence related to the distribution, sale, acquisition, and use of thousands of surviving 15$^{th}$-century printed books.

\subsection{Data}
\label{sec:Sec2.3}
\begin{figure}[t!]
 \centering
 \includegraphics[width=\columnwidth]{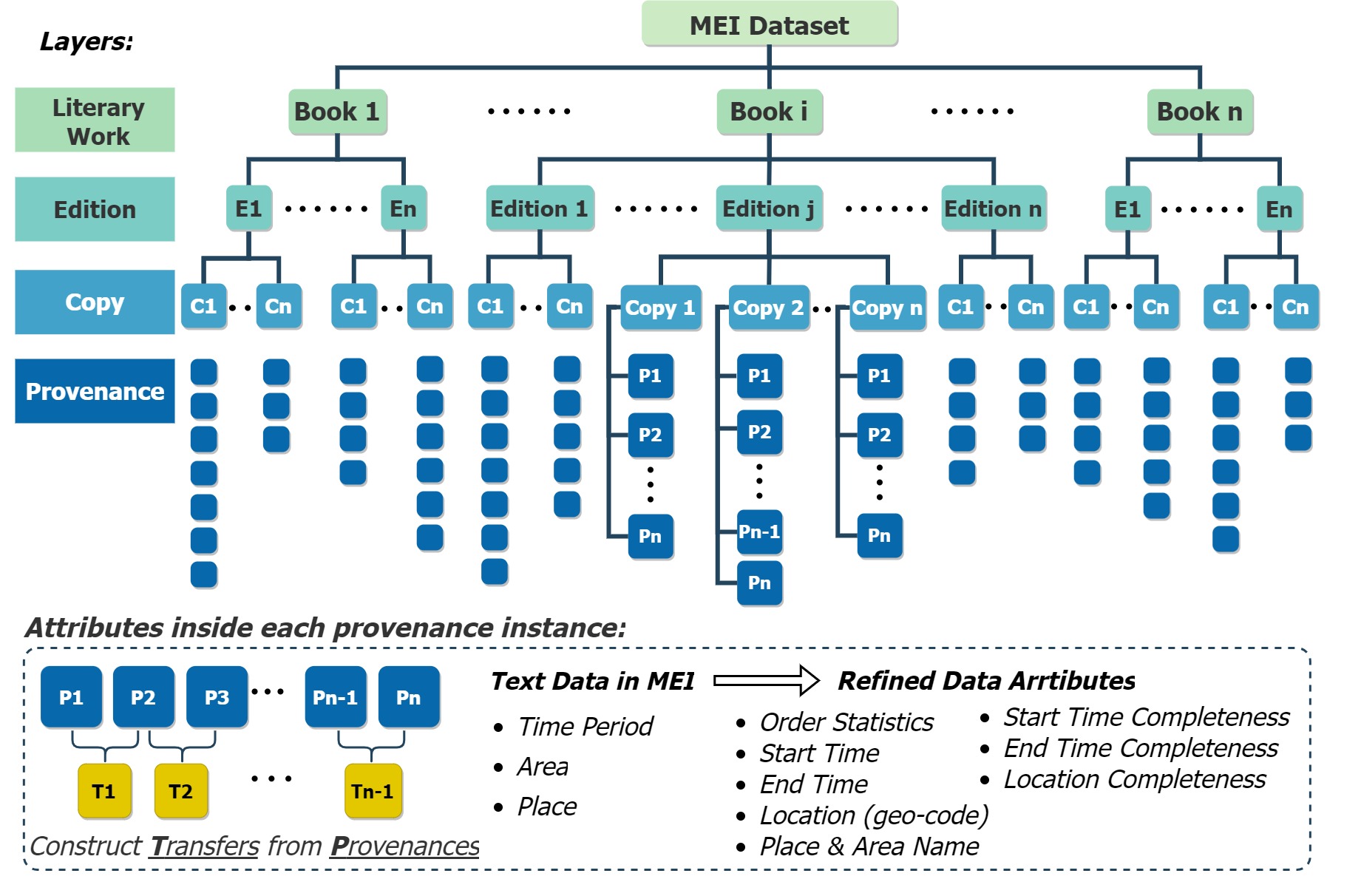}
 \caption{The hierarchical structure of MEI data (top) and the evolution of attributes in provenance instances (bottom).}
 \label{fig: nested data structure}
 \vspace{-0.15in}
\end{figure}

Although MEI records are presented in a cataloged format, they are inherently  hierarchical (Fig.~\ref{fig: nested data structure}). A specific work of literature can have multiple print editions,  distinguished by its unique ISTC (Incunabula Short Title Catalogue) codes. Each print edition may have multiple copies, a copy is assigned a unique MEI ID, and its physical evidence is recorded chronologically through a series of provenance blocks.
A set of records entered into MEI for the Polonsky Foundation Dante Project was used in \emph{DanteExploreVis}.
An illustrated copy census of all the 173 surviving copies of the first Florentine edition of Dante's Commedia, printed in 1481 showed that these copies are scattered worldwide. 
This edition was chosen because it contains the complete copy census on the distribution, use, and survival of copies of an edition.

\subsection{Previous Work}
Prior to initiating the design study on \emph{DanteExploreVis}, a separate design study on the same dataset, addressing different domain problems, was carried out following the `Nine-Stage Framework'. This process entailed a close collaboration with the same leading domain expert and led to the development of \emph{DanteSearchVis}, the inaugural visualization tool under the \emph{BookTracker} platform for visualizing MEI data. The tool was subsequently tested and evaluated by five additional domain experts for its usability and effectiveness.
The insights and reflections gleaned from the previous design study were summarized~\cite{xing2022design}, prompting an adjustment of the framework to better suit our needs.  
Communication with the leading domain expert persisted following the deployment of \emph{DanteSearchVis}. Feedback and expectations from the five domain experts involved in the evaluation were extensively discussed with the leading domain expert. Both parties concurred on the merit of further visualization research using MEI data and anticipated incorporating new features into the \emph{BookTracker} platform to fulfill diverse domain requirements. The cross-disciplinary collaboration demonstrated its value and ongoing relevance, warranting continued exploration.
\section{Related Work}
\label{sec:Sec3}

\subsection{Visualization Design Study}
Core to visualization design studies is a collaboration between researchers and domain experts to address real-world problems using visualization techniques. The process includes developing a solution to the problem and evaluating it in retrospect.
A rich literature exists on models and methodologies for conducting design studies. 
The Nested Model~\cite{munzner2009nested} provides prescriptive guidance for visualization design and validation by organizing the process into four cascading levels. The Nine-Stage Framework~\cite{sedlmair2012design} provides a structured approach with 9 stages to conduct problem-driven visualization research by working with domain experts. Each stage is designed to ensure that the resulting visualizations meet the needs of stakeholders and are appropriately validated at each level of the process. These two methodologies are widely applied to visualization design studies for their generalizability, flexibility, and adaptability~\cite{abdul2014comparing, zhang2018idmvis, nobre2018lineage, brehmer2014overview, eirich2021irvine, ye2020user}. 
The Design Activity Framework~\cite{mckenna2014design} breaks down the design process into four series of design activities and linked them to the four layers of the Nested Model, which offers a more flexible structure for iteration. The Design Study `Lite' Methodology~\cite{syeda2020design} is an expedited framework for visualization design research within limited time frames. The notion of the data-first design study and refinement of the Nine-Stage Framework to fit the data-driven visualization research were proposed in~\cite{oppermann2020data}.
Our review, of papers on visualization projects in the field of humanities, found commonalities in the practice of these projects~\cite{arnold2020visualizing, ciula2021small, zhang2022cohortva, muller2021uncertainty, vancisin2020externalizing}. Combining our own experience working with historical book researchers~\cite{xing2022design}, we consider that the core stages in the Nine-Stage Framework can be adjusted and refined to be more suited and adaptable to the design studies in collaboration with humanities researchers, particularly in terms of improving understanding and communication and obtaining more comprehensive data and problem abstractions. 
\rev{In the domain of Human-Computer Interaction (HCI), iterative design principles~\cite{buxton2010sketching, camburn2017design} and user-centered design methodologies~\cite{da2011user, abras2004user, BRHEL2015163} significantly inform our design study, which places domain experts at its core. By drawing from the widely acknowledged design lifecycle~\cite{zhang2005integrating,mayhew1999usability} in HCI and User Experience (UX) realms, we tailored and melded elements such as need-finding, design alternatives, prototyping, and validation into the Nine-Stage Framework~\cite{sedlmair2012design}. This resulted in a more manifest representation of the iterative design nature subtly embedded in the core stage of the Nine-Stage Framework, thereby bolstering its usability and actionability across a diverse range of domains.}
To ensure the rigor of the evaluation practice in our study, we have referred to the following evaluation-related works for theoretical support: 1) evaluation on human-centered processes~\cite{huang2014handbook, tory2005evaluating}, 2) qualitative evaluation methods~\cite{fonteyn1993description, dingman2021interview, forsell2012guide}, and 3) usability evaluation~\cite{nielsen1994usability, winter2021exploring, brinck2001usability}.
\subsection{Map-Based Visualizations on Trajectory Data}
Consisting mainly of ordered provenance (OD) information, the book trade data is a sequence of spatial points arranged according to timestamps and carry additional textual information. Although it is atypical due to its sparse time points and irregular intervals, it can be considered as trajectory data.
To gain inspiration for our work, we have perused papers on map-based visualization of trajectory data.
Dynamic vehicle movement tracing~\cite{sobral2019visualization, huang2015trajgraph} and population mobility~\cite{redin2017vitflow, skupin2005visualizing, ling2016classifying} are hot topics under the umbrella of visualizing trajectories. In a survey paper by He et al.~\cite{he2019variable}, various visual representations and techniques for multivariate spatio-temporal trajectory have been compared. 
Heatmaps~\cite{guo2007visual, wood2010visualisation} are commonly used to represent OD data to show the event volume and density, while force-directed edge-bundling~\cite{holten2009force} is used as a bundling technique to reduce visual clutter on the map.
\subsection{Book Trade Related Projects}
Previous research that shared common interests with historical book analysis was reviewed. Both Peripleo~\cite{simon2016peripleo} and ArtVis~\cite{dumas2014artvis} displayed spatio-temporal historical data using scatterplot-based visualizations. However, they did not depict sequential paths. Visualizations of the Republic of Letters~\cite{chang2009visualizing, edelstein2017historical} succeeded in illustrating correspondence circulation, but their primary focus was on representing the overall volume of letter exchanges rather than the narrative of an individual record.
Regarding visualizations of the book trade data, `The Atlas of Early Printing'~\cite{Atlas:2021:web} utilized a GIS map of Europe to illustrate the spread of printing and typography, trade routes, and other historical data. The `MEI Map Current'~\cite{cerl_mei_map} displayed the distribution of the location where the books are being held today and was linked to the MEI database. 
Currently, no tools concentrate both on distribution and movements of books, therefore, this becomes the emphasis of our work.
\section{Design Study Overview}
\label{sec:Sec4}
Collaborating with historians highlights unique aspects, including the need for clear communication of academic terminology, the potential for humanities researchers' divergent thinking to affect project direction, and the evolving data perceptions when using visualization tools may present new research opportunities. Although the Nine-Stage Framework serves as a valuable design study methodology and offers general guidance, its highly structured and repetitive reflective loop may be inefficient or lack direction in specific contexts. Consequently, we recommend adapting its \textit{core} phase with iterative design cycles to ensure thorough evaluation and reflection while maintaining consistent interactions with humanities experts.
We refer to Leading Domain Expert as ``\textbf{LDE}'' and larger Domain Experts' group as ``\textbf{DEs}''.

\subsection{Variant of the Nine-Stage Framework}
\rev{The adapted framework maintains the \textit{precondition} and \textit{analysis} phases in the Nine-Stage Framework while restructuring the \textit{core} into iterative cycles with modified objectives.
We tailored the \textit{core} phase (\textit{discover}, \textit{design}, \textit{implement}, \textit{deploy}) to our context, retaining the first 3 stages in the development cycles with weekly meetings as iteration nodes. We included \textit{validate} in each iteration for controllable evaluation and regular communication with DEs.} 

\textbf{Discover:} 
In addressing the challenge of an early \textit{Discover} stage for problem abstraction~\cite{sedlmair2012design}, our adapted framework places \textit{Discover} at the beginning of each iteration cycle to: 1) refine domain problems, 2) increase DEs' interaction 
and 3) expand information collection channels beyond the conversation. We observed that DEs' ability to ask pertinent questions depends on their data understanding and familiarity with data. For example, historians with limited quantitative analysis experience and reliance on textual data may struggle to identify deeper concerns as data volume grows. We moved data abstraction from the \textit{design} to the \textit{discovery} stage, enabling DEs to improve their domain data knowledge with 
assistance and visualization tools at each iteration's onset. 

\textbf{Design:} 
In the adapted framework's iteration cycle, we revised the \textit{design} stage, emphasizing task abstraction, and visual and interaction design. We integrate task abstraction approaches~\cite{brehmer2016why,brehmer2014overview,munzner2014visualization} along with the Nine-Stage Framework. DEs often have specific design expectations (e.g., our LDE desired a 2D geopolitical map with all elements
); we recommend first summarizing their anticipated visualizations and then adjusting or proposing alternatives from visualization experts' perspectives. Presenting them with multiple design options using real data to test and then implement the highest-rated design in the final tool.

\textbf{Implement:} In the adapted framework, we suggest using rapid prototyping with real data to explore the design and refine the early versions of the tools. As iterations are marked by regular meetings with the LDE, we recommend parallel prototyping within each iteration, simultaneously presenting design alternatives for various components for evaluation in the subsequent validation stage.

\textbf{Validate \& Evaluation:} 
In the adapted framework, we split user evaluation into: 1) informal evaluations during discussions with the LDE, and 2) systematic evaluations involving a larger group of DEs prior to deployment. The former, termed \textit{validate}, is incorporated within the iteration cycle to assess the usefulness of features in early versions. We asked questions such as \say{Do the current visualizations address domain requirements?} and \say{Which design alternatives are most effective?} during validation. Usability feedback is collected but not prioritized. The latter, labeled \textit{evaluation}, occurs after the final iteration, outside the iteration cycle, and before the tool deployment. This comprehensive evaluation focuses on usability and usefulness, engaging a broader user group to ensure unbiased results.

\rev{Taking cues from established iterative prototyping and user-centered design strategies common in HCI, we modified the Nine-Stage Framework to accentuate the importance of explicit iteration in its \textit{core} phase, and the necessity for significant involvement of domain experts, especially during domain data, problem, and task abstraction stages. Although the iterative essence exists subtly in the original framework, we have given it a more explicit prominence in our adaptation, thereby amplifying its actionability and extensive applicability. This enhancement strengthens the framework's systematic compliance and usability across multiple domains, signifying its domain-agnostic qualities.}

\subsection{Design Study Following the Framework Variant }
Before initiating the design study for \emph{DanteExploreVis}, we already had collaborations with researchers in historical books to develop \emph{BookTracker}. The established workflow allowed us to swiftly progress through the \textit{precondition} phase.
The design study began in May 2022 and lasted six months. In the first two weeks, we clarified domain problems and data sources and set a timeline with the LDE. Between June and August, we completed seven design iterations, resulting in \emph{DanteExploreVis}. Subsequently, a two-month evaluation period included surveys and expert interviews to assess the tool's usability and usefulness. Finally, we documented our findings and reflections in the analysis phase, presented in this paper. Fig.~\ref{fig:teaser} illustrates the timeline, interface evolution, and key issues discussed in each iteration.
\section{Domain Problems and Tasks}
\label{sec:Sec5}
In this section, we present abstractions from the design study, encompassing requirements, data, and tasks. We initially identified six domain requirements and captured evolving requirements during the design process. We also outline five essential data modifications and additions. Furthermore, we analyze each domain requirement, both original and derived, associating them with low-level tasks.
%
%
\subsection{Problem Abstraction and Requirement Analysis}
In the \textit{precondition} stage, the LDE described their research objectives and 
the historians' research flow using MEI. She emphasized that users rely on MEI to formulate initial queries related to the early distribution, later survival, and circulation of knowledge in 15$^{th}$-century printed books. They also sought to uncover patterns of book ownership over time, such as institutions or individuals as well as entangled histories and shared heritage, revealed through book ownership.
MEI provides basic query and display features, allowing users to retrieve book editions using ISTC numbers. The catalog-style listing and separate pages for textual provenance information are suitable for close reading but lack overview and exploratory analysis. The temporal and spatial aspects of the provenance data are not well presented too. In response to visualization needs, together with the LDE, we documented the problems that could not be directly addressed by MEI:
%
\begin{itemize}[noitemsep,leftmargin=*]
    \item Search for books with similar features: \say{\textit{Which copies were printed in Italy and used in Germany?}}
    \item Visualize the circulation of one edition intuitively: \say{\textit{Can I see the circulation of all surviving copies with their provenance history of Dante’s Commedia printed in \rev{Florence} in 1481?}}
    \item Insights from the provenance: \say{\textit{Can we extract more insightful summaries from the valuable provenance data in MEI?}}
\end{itemize}
From communications with the DEs, we distilled six main requirements and listed them along with derived requirements from iteration cycles.
\begin{description}[noitemsep,nolistsep,leftmargin=0.05in]
    \item[R1] \textbf{Visualize the trajectory of a single copy.} 
    The data contains valuable spatio-temporal features not represented in the MEI interface. DEs anticipate that the visualization tool will intuitively display the transfer trajectory of each copy. During the design process, three additional requirements were derived:
    \begin{description}[noitemsep,nolistsep]
        \item [R1.1] \textit{Show an overview of full journey and provenance breakdown.}
        \item [R1.2] \textit{Animate the provenance-by-provenance movement.}
        \item [R1.3] \textit{Establish connection to the MEI page of each copy.}
    \end{description}
    
    \item[R2] \textbf{Visualize the circulation of a group of copies.}
    Copies with shared geo-features are often of research significance, such as those that have been transferred from Italy to the United States, potentially indicating important historical events in the printing revolution. Since MEI can only list individual copies, the visualization tool should display groups of copies with common characteristics, such as having similar transfer experiences or being stored at the same location, in a single view for deeper insights. Derived requirements from this one include:
    \begin{description}[noitemsep,nolistsep]
        \item[R2.1] \textit{Enable users to identify copies with shared characteristics.}
        \item[R2.2] \textit{Offer insights into the movement of groups of copies.}
    \end{description}
    
    \item[R3] \textbf{Present the distribution and circulation of all copies on a geopolitical map.}
    The circulation of an edition is formed by the distribution of all its copies. Understanding the geographical distribution of the edition throughout its circulation can help shed light on historical issues. 
    The DEs stress the significance of visualizing the trajectories of all copies on a geopolitical map. The derived requirements include:
    \begin{description}[noitemsep,nolistsep]
        \item[R3.1] \textit{Gaining insights on path density and clustering.}
        \item[R3.2] \textit{Capability to overview individual instances.}
    \end{description}

    \item[R4] \textbf{Static visualization of provenance-related information.}
    The primary value of the MEI lies in the reconstruction of provenance, which is presented in a limited way by the MEI interface. An overview of all crucial aspects of the provenance instances, such as the time period and location of stay, for each copy was desired. Simultaneously, due to their evolving comprehension of the data, they wish to indicate the data completeness of elements within each provenance, facilitating future data entry and revision efforts. Derived requirements include:
    \begin{description}[noitemsep,nolistsep]
        \item[R4.1] \textit{Visualize \rev{data uncertainty (completeness)} for each provenance.}
        \item[R4.2] \textit{Provide general statistics on number of provenances.}
    \end{description}
    
    \item[R5] \textbf{Animate book movement.}
    DEs wish to view book transfers in animation, particularly for presentation and enjoyment purposes. They find that animated presentations facilitate their understanding of event chronology. A requirement associated with animation is:
    \begin{description}[noitemsep,nolistsep]
        \item[R5.1] \textit{Provide different ways to play the animation.} 
    \end{description}
        
    \item[R6] \textbf{Gain insights from features extracted from provenance-related information.}
    DEs take pride in the reconstruction of provenance for each copy and believe that the data can offer far more insight than the current MEI interface.
    They hope the visualization tool can characterize the dataset at a higher level of abstraction and enable them to examine the dataset from various perspectives. Two requirements are raised:
    \begin{description}[noitemsep,nolistsep]
        \item[R6.1] \textit{Examine the dataset from different angles.}
        \item[R6.2] \textit{Expect extracted features to be used as query entries.}
    \end{description}    
\end{description}
\subsection{Data Abstraction}

\begin{figure}[t!]
 \centering
 \includegraphics[width=\columnwidth]{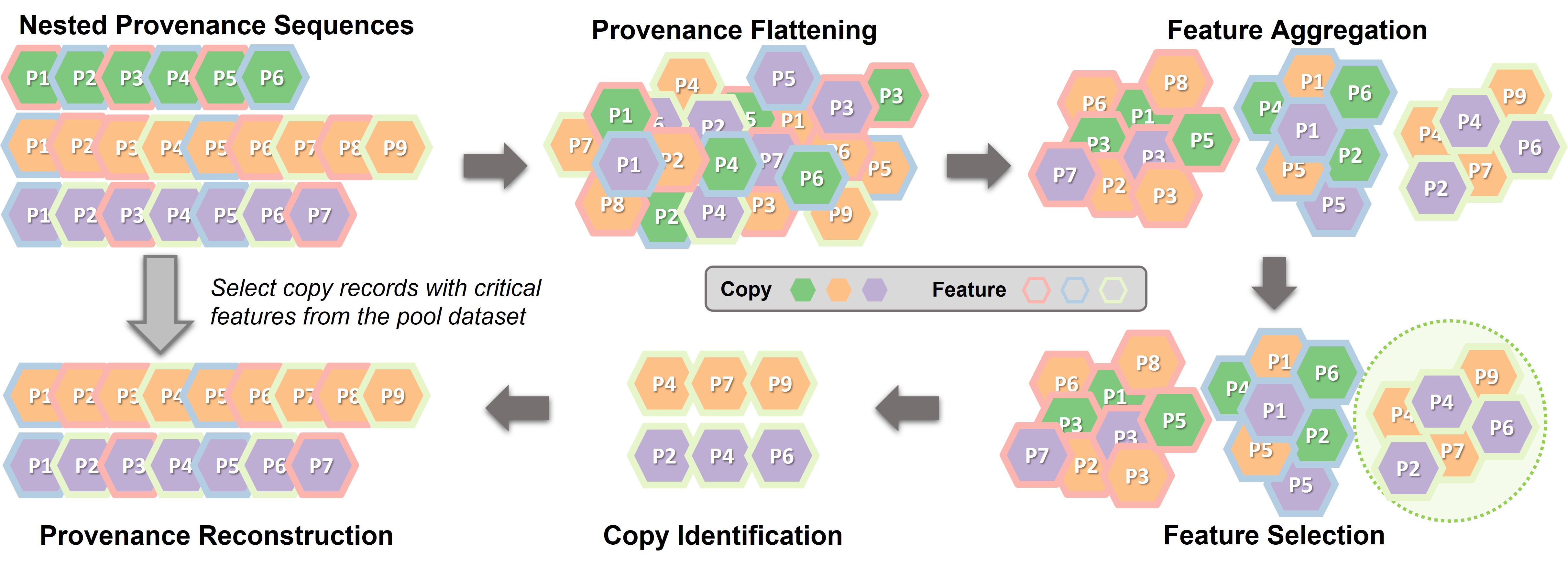}
 \caption{Encode the data to obtain the subset with critical features.}
 \label{fig:data encoding}
 \vspace{-0.15in}
\end{figure}
The MEI data (in Fig.~\ref{fig: nested data structure} (top)) can be interpreted as a hierarchical structure. The outermost layer is the database, which stores numerous literary works as \textbf{\textit{book}} nodes. \textbf{\textit{Editions}} throughout history become sub-nodes, identified by ISTC codes. Each edition produces printed \textbf{\textit{copies}}, assigned unique MEI IDs, serving as the lowest-level nodes. Ordered \textbf{\textit{provenance}} instances in each copy, which DEs seek to visualize and analyze, differentiate the MEI data from other historical book datasets. We divided the textual data inside each provenance instance into two categories: conclusive data (\textit{time and location data used to present the status of the provenance}) and evidential data (\textit{marginal annotations and binding styles, etc.}) for obtaining and confirming conclusive data. From discussions with the LDE, geographical and chronological information related to the book's provenance was prioritized for visualization.

Adhering to the framework variant, we incorporate data discovery at the beginning of each iteration, continuously performing data abstraction throughout the design study, and seizing every opportunity for DEs to reflect on the data and generate new requirements. We expanded the existing MEI data to better address domain needs. We outline four manipulations made during the iterations, which can be regarded as milestones propelling the progress of the design study:
%
\begin{description}[noitemsep,nolistsep,leftmargin=0.05in]
    \item[D1] \textbf{Addition of Geographical Coordinates:} 
    Building on our past collaboration with the DEs, we reached a consensus that the tool should concentrate on displaying book transfer paths on a geopolitical map. Following the acquisition of raw JSON data from the MEI database, our initial data processing step entailed converting all area and place names into corresponding geographical latitude and longitude coordinates, facilitating their visualization on the map.

    \item[D2]\textbf{Alternative Presentation of Chronology:} 
    The LDE emphasized MEI's uniqueness in aggregating and integrating each book's provenance across different time periods. Provenance instances are linked chronologically, forming the book's historical circulation trajectory. In addition to displaying each provenance location on the map, we aimed to visualize the transfer path, direction, and sequence. From the raw data, we obtained each provenance's time period (in years). Ideally, the current transfer's start and end times should match the end time of the current provenance and the start time of the next one. We separated each provenance's time period into start and end times, then reconstructed them to obtain each transfer's start and end times. The symbolic representation is as follows:

    For the $i^{th}$ copy $C_i$, let $P_i=\{p_1,p_2,....,p_n\}$ be the set of provenance instances, where $n$ is their total number. $T_i=\{t_1,t_2,...,t_{n-1}\}$ denotes transfers between consecutive provenance instances. For the $j^{th}$ provenance, the time range is $[p^j_{start},p^j_{end}]$, so the start and end time pairs for the $j^{th}$ transfer can be denoted as $[t^j_{start},t^j_{end}] = [p^j_{end},p^{j+1}_{start}]$.
    %
    After splitting the time periods, we encountered issues constructing a coherent time series for book transfers and stays. Transfers with end time earlier than start time ($t^j_{end}<t^j_{start}$) indicated numerous provenance instances in the original data with $p^{j+1}_{start}<p^j_{end}$. The main reason was the imprecision of provenance time periods in MEI data. We identified two main data sources: 1) precisely recorded data from libraries or purchase records and 2) inferences by historians based on material evidence. After consulting with the LDE, we agreed that improving accuracy would be difficult. Instead, we decided to use the order statistic $j$, as the relative time of transfer occurrence.

    \item[D3] \textbf{Addrressing Data \rev{Uncertainty}:}
    We discovered that ambiguity exists not only in temporal features but also in geographical information for each provenance instance. Similarly to the temporal aspect, some provenance instances lack clearly documented locations, and historians may or may not estimate the location roughly based on existing material evidence. \rev{Here, we introduced three additional attributes using three data completeness levels to indicate the uncertainty of \textit{start time}, \textit{end time}, and \textit{location}}. 
    The completeness levels are: 
    1) \textbf{\textit{Accurate}} -- Clear recorded information is available;
    2) \textbf{\textit{Approximate}} -- Estimation based on the material evidence;
    3) \textbf{\textit{Missing}} -- No obvious evidence or information is available.

    \item[D4] \textbf{Flatting Provenance Blocks:}
    To address the limitations of displaying provenance information in the sequenced blocks confined in a hierarchical structure, as they obscure non-temporal features. We proposed a de-hierarchized and de-ordered approach (Fig.~\ref{fig:data encoding}) allowing DEs to analyze each provenance or transfer event individually. The deconstructed data is clustered based on shared attributes, such as transfer location, and subsequently reorganized to create a subset of MEI records conforming to specific criteria.

    \item[D5] \textbf{Bundled Path Coordinates:}
    In designing \textit{Single-Static Storyboard} (Fig.~\ref{fig:DanteXploreVis} d2), we faced the challenge of visual clutter when displaying all paths as straight lines on the map. To present all paths while minimizing visual clutter, we implemented the edge bundling approach and employed the force-directed edge bundling algorithm~\cite{holten2009force} to generate a list of coordinates for curved lines, effectively replacing the initial straight lines for each transfer trajectory.
\end{description}
%
%



%
\subsection{Task Abstraction}
Task abstraction follows behind requirement and data abstraction while maintaining a strong connection to subsequent designs. In line with the adapted framework, we conducted task analysis during each design iteration, particularly when new discoveries emerged. 
We identified three main objectives derived from domain requirements: 
1) \textbf{Explanatory Analysis:} 
\rev{presenting data directly and intuitively, enabling users to uncover the story behind each copy and provenance record;}
2) \textbf{Exploratory Analysis:} 
\rev{exploring data to gain insights and deeper understanding through visualizations;}
3) \textbf{Presenting \& Enjoying:} 
\rev{displaying data for audiences or deriving pleasure from visualization results.}
%

Drawing from the approach to task abstraction in~\cite{brehmer2016why,6634168,munzner2014visualization}, we further refined the high-level domain problems into lower-level tasks based on the purpose of each requirement. \textbf{T1 \& T2} stemmed from the need for exploratory and explanatory analysis, respectively. \rev{\textbf{T3 \& T4} were intermediate-level tasks for in-depth exploration and explanation of the data of interest after the initial exploration phase, while \textbf{T5} addressed the presenting and enjoying purpose.}
The set of design tasks crafted to fulfill the domain requirements includes:
\begin{description}[noitemsep,nolistsep,leftmargin=0.05in]
    \item[T1] \textbf{Browse \& Explore:} 
    For  \textbf{R1.1, R2.2, R3, R3.1, R4, R4.1, R4.2, R6, R6.1}.
    The MEI interface provides extensive textual information for each provenance, but lacks a way to browse the entire dataset at a glance. Offering multiple exploration methods through various entry points and data layers can reveal diverse insights and is crucial.
    \item[T2] \textbf{Elaborate \& Explain:}
    For \textbf{R1, R2.}
    The textual data in MEI are insufficient to present data with temporal and spatial features. Map-based visualizations are necessary to elaborate and interpret the data.
    \item[T3] \textbf{Lookup \& Locate:}
    For \textbf{R1.3, R6.2}.
    When searching for a specific target, such as a copy with a known ID, 
    the tool should offer functionalities to look up and locate the target for further exploration.
    \item[T4] \textbf{\rev{Identify \& Compare:}}
    \rev{For \textbf{R2, R3}.
    When examining multiple copies, the tool should enable users to identify historical phenomena and points of interest, and to compare the movements of different copies, observing similarities and dissimilarities.}
    \item[T5] \textbf{Present \& Enjoy:}
    For \textbf{R1.2, R3.2,  R5, R5.1}.
    To enhance the audience's perception and sense of immersion in terms of presentation and enjoyment, the tool needs to add animation to the static visualization of the book's movement. 
\end{description}

\section{Iterative Design Process}
\label{sec:Sec6}
In this section, we \rev{briefly} discuss the iterative design process in accordance with the framework variant, providing an explanation for the timeline shown in Fig.~\ref{fig:teaser}, while emphasizing the iterative evaluation-based improvement. \rev{A more detailed account is available in the supplementary material.}

\textbf{Precondition: \textit{30 May -- 23 Jun}.} 
Potential developments for visualizing the MEI data were discussed in an initial meeting with the LDE. We identified four unmet requirements: 1) presenting the trajectory of each copy individually, 2) visualizing and comparing the circulation of multiple copies simultaneously, 3) exploring and gaining insights from provenances, and 4) implementing animations to enhance the understanding of provenance sequence. 

\textbf{Iteration 1: \textit{23 Jun -- 05 Jul}.}

\noindent \underline{\textbf{\textit{Discover:}}}
Grounded in domain challenges and expectations from the \textit{precondition} phase, six domain requirements (\textbf{R1}-\textbf{R6}) were established. Geocoding locations (\textbf{D1}) and splitting provenance periods were discussed (\textbf{D2}). However, data \rev{uncertainty} posed challenges.

\noindent \underline{\textbf{\textit{\rev{Design \& Implement}:}}}
Focusing on explanatory and exploratory analysis of provenance data, domain requirements were transformed into lower-level visualization tasks. Three solutions for visualizing ordered paths on a map were proposed: 1) multiple views showing path development over time, 2) gradient color rendering of paths, and 3) path animation. For data exploration, feature aggregation was suggested.
%
\rev{Designs ideas were documented in Fig.~\ref{fig:teaser} I1, emphasizing ideation.}

\noindent \underline{\textbf{\textit{Validate:}}}
Ideas were discussed with the LDE, who found all path visualization methods useful in different contexts. The expert questioned the feature aggregation proposal, but was open to seeing a more concrete demo before deciding. Data \rev{uncertainty} was acknowledged as inevitable and the reasons behind it were explained.

\textbf{Iteration 2: \textit{05 Jul -- 26 Jul}.}

\noindent \underline{\textbf{\textit{Discover:}}}
Focused on providing an overview and insights (\textbf{R6}), this iteration proposed simplifying trajectory data to origin-destination data and handling incomplete geographical and temporal information. Order statistics were suggested as an alternative to showing sequence (\textbf{D2}).

\noindent \underline{\textbf{\textit{\rev{Design \& Implement}:}}}
The interface was tentatively divided into two parts. The left view provided exploratory analysis using heatmaps, while the right offered explanatory analysis based on geopolitical maps.
%
A simple, real data-based demo was implemented (Fig.~\ref{fig:teaser} I2 \& I3).

\noindent \underline{\textbf{\textit{Validate:}}}
The LDE found the heatmap design for feature aggregation more useful than anticipated, it was comprehensible and effectively emphasized data traits. She acknowledged the value of breaking down the provenance (\textbf{D4}) and the possibility of presenting them in heatmaps to identify the location and temporal characteristics. She agreed with the redefinition of transfer and the use of order statistics instead of the year (\textbf{D2}), pressing the need for a full presentation of geographical locations and provenance occurrence order.

\textbf{Iteration 3: \textit{26 Jul -- 01 Aug}.}

\noindent \underline{\textbf{\textit{Discover:}}}
Different levels of data completeness were discussed (\textbf{D3}). The refined data structure with additional attributes is shown in Fig.~\ref{fig: nested data structure} (bottom). The LDE's interest in provenance data completeness (uncertainty) resulted in new domain requirements (\textbf{R4.1} \& \textbf{R4.2}). Regarding exploratory visualizations, \textbf{6.1} was formulated. To visualize individual copy trajectories (\textbf{R1}), we expanded the requirements based on the LDE's suggestions, adding \textbf{R1.1}, \textbf{R1.2}, and \textbf{R1.3}.

\noindent \underline{\textbf{\textit{\rev{Design \& Implement}:}}}
To address requirements \textbf{R1}, \textbf{R2}, and \textbf{R6}, we proposed three explanatory storyboards to cater to different requirements: 1) \textbf{\textit{Multi-Static Storyboard}} \textit{(or \textbf{MSS})} for individual provenance through multiple concatenated static views, 2) \textbf{\textit{Single-Static Storyboard (SSS)}} for a single static view of multiple copy trajectories, and 3) \textbf{\textit{Single-Dynamic Storyboard (SDS)}} for presenting the journey of multiple copies in animation.
%
For the \textbf{\textit{MSS}}, an overview and step-by-step sequence visualizations were implemented for each copy. The overview included a radar chart to display data completeness and a map to demonstrate the complete trajectory of a copy. Two additional heatmaps were implemented to satisfy \textbf{R6.1}.

\noindent \underline{\textbf{\textit{Validate:}}}
The LDE found the \textbf{\textit{MSS}} to be valuable and appreciated both the overview and step-by-step sequence visualizations. We were suggested to improve the design by connecting the overview and step-by-step sequence maps, allowing users to easily pinpoint meaningful provenances. She also expressed that the radar chart can effectively display provenance information and data \rev{uncertainty}, but requested the addition of \rev{uncertainty} hints in the step-by-step sequence map.

\textbf{Iteration 4: \textit{01 Aug -- 08 Aug}.}

\noindent \underline{\textbf{\textit{Discover:}}} 
The \textbf{\textit{MSS}} was well-received but needed improvements in interactivity and view associations. Key issues to address included overlapping provenances, chronological representation, associations between overview and step-by-step maps, MEI database linking, highlighting similar transfer instances, and adding textual descriptions.

\noindent \underline{\textbf{\textit{\rev{Design \& Implement}:}}}
To better illustrate the path sequence, several designs were proposed: 1) 
\rev{a conical gradient circle glyph indicates total provenances and current timeline position.}
2) a secondary view showing a horizontal journey of provenances over time, and 3) add animations to the overview map. 
%
We implemented the proposed designs, including a horizontal journey with clickable circles connected to corresponding provenances in the sequenced maps, and emphasized line segments for identifying and comparing similar transfers. The animation was added to the overview map, and donut glyphs displayed the \rev{data uncertainty} on the sequenced maps. Links to MEI record pages were also provided.

\noindent \underline{\textbf{\textit{Validate:}}}
The LDE appreciated the horizontal journey visualization and the direct link to the individual MEI record pages. Animations enhanced the transfer visualization experience. The expert preferred simpler monochrome fills for points on the overview map but suggested retaining both designs (Fig.~\ref{fig:teaser} S4) for user preference testing. With a single-copy presentation exceeding expectations, the focus shifted to visualizing multiple copies (\textbf{R2} and \textbf{R3}).

\textbf{Iteration 5: \textit{08 Aug -- 15 Aug}.}

\noindent \underline{\textbf{\textit{Discover:}}}
Regarding presenting multiple trajectories simultaneously (\textbf{R2} \& \textbf{R3}), additional domain requirements were identified by the LDE: visualizing distribution and aggregation patterns (\textbf{R3.1}), and identifying trajectories from the map (\textbf{R3.2}). These requirements were associated with the development of \textbf{\textit{SSS}}. However, displaying all paths on a single map led to visual clutter, so we incorporated edge bundling to improve path aggregation and minimize clutter (\textbf{D4}).

\noindent \underline{\textbf{\textit{\rev{Design \& Implement}:}}}
Bundled paths were proposed to decrease visual clutter and emphasize clustering, but the degree of bundling needed to be evaluated. To identify single copies, we designed interactions with the \textbf{\textit{information panel}}, allowing users to highlight and compare multiple trajectories.
%
We implemented these features. 
\rev{We generated five visualizations using edge bundling with different parameters ranging from minimum to maximum, for the LDE to test (Fig.~\ref{fig:teaser} S5).}

\noindent \underline{\textbf{\textit{Validate:}}}
The LDE valued the \textbf{\textit{SSS}} and its ability to highlight individual or multiple copies. Initially, she favored straight paths over bundled curved ones, but as she interacted with the prototype, she began to appreciate the advantages of bundling in displaying aggregates. Ultimately, we chose to retain both straight and bundled paths, leaving the bundling degree for broader evaluation.

\textbf{Iteration 6: \textit{15 Aug -- 22 Aug}.}

\noindent \underline{\textbf{\textit{Discover:}}} 
With most visualization features implemented, the LDE focused on more practical domain needs -- the query methods. The heatmap, while informative, was not familiar to DEs for searching and querying. Requirements for other query methods, such as multi-entry and single-entry searches, were mentioned (\textbf{R2.1 \& R2.2}).

\noindent \underline{\textbf{\textit{\rev{Design \& Implement}:}}}
Following discussions on DEs' familiar search methods and specific issues, we incorporated two designs into the query section \rev{as separate tabs}: 1) searching by setting print and current locations; and 2) searching by individual MEI ID. The drop-down menus are designed to input query constraints.
%
The initial \textbf{\textit{SDS}} was developed, with moving markers showing book transfers.

\noindent \underline{\textbf{\textit{Validate:}}}
The LDE appreciated the two new convenient search methods with drop-down inputs and was delighted by the animated transfer histories on a single map. She then recommended two distinct playback sequences to display the animation of multiple copy transfers.

\textbf{Iteration 7: \textit{22 Aug -- 29 Aug}.}

\noindent \underline{\textbf{\textit{Discover:}}}
We further refined animation-related designs after discussing them with the LDE. She suggested two ways to present multiple trajectories: simultaneously or sequentially. We labeled this as new derived requirements \textbf{R5.1}. Furthermore, we needed to address the identification issue for multiple paths in the \textbf{\textit{SDS}}.

\noindent \underline{\textbf{\textit{\rev{Design \& Implement}:}}}
As requested, we designed and \rev{implemented} a sequential playing mode in which only one moving marker appears at a time, while completed trajectories stay on the map in different colors. To address the issue of identifying individual copies, we proposed and implemented clickable moving markers that display copy IDs and provide links to their respective MEI pages.
%

\noindent \underline{\textbf{\textit{Validate:}}}
All requirements and derived needs were met, and the LDE was satisfied with the tool, suggesting no further improvements.

\textbf{Evaluation \& Deployment: \textit{29 Aug -- 29 Nov}.}
The tool was further evaluated by a larger group of DEs to mitigate the potential bias of the LDE, as detailed in Section~\ref{sec:Sec8.1}.

\textbf{Analysis: \textit{29  Nov -- 31 Dec}.}
After the evaluation and deployment, the outcomes were discussed with the LDE. 
The design study concluded with a summary of crucial aspects and lessons learned.
%
\section{DanteExploreVis}
\label{sec:Sec7}

\begin{figure*}[ht]
 \centering
 \includegraphics[width=0.95\textwidth]{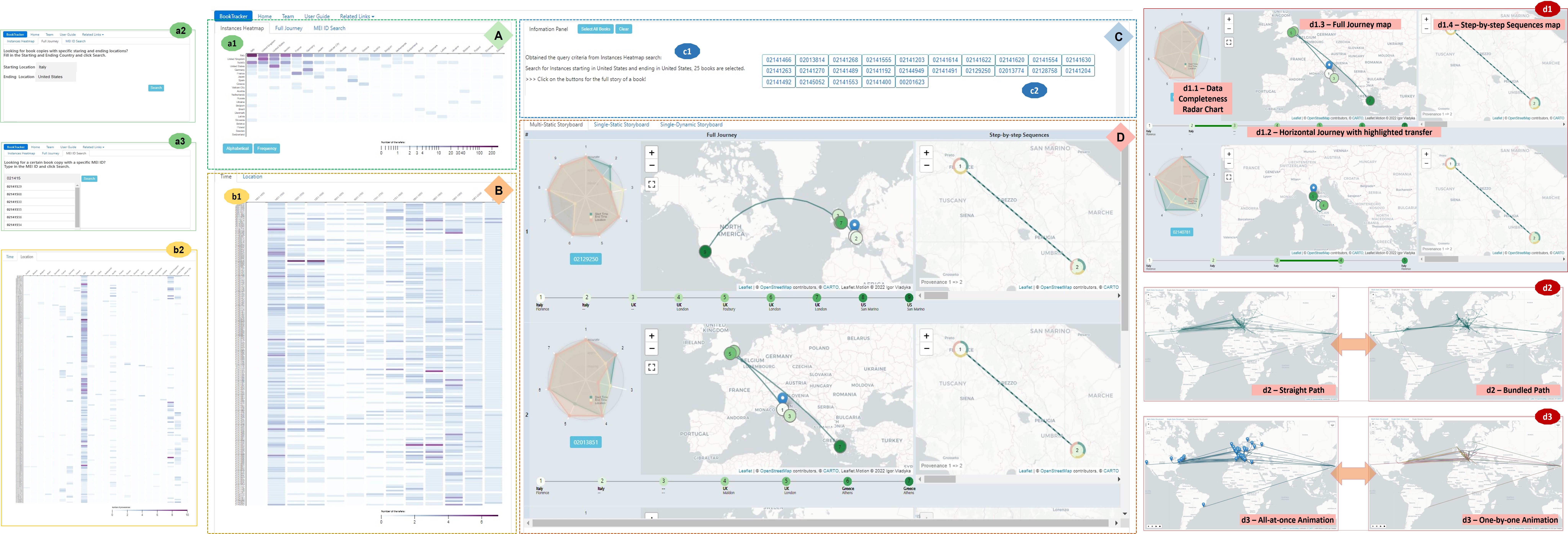}
 \caption{The interface of DanteExploreVis, highlighting its key components: (A) and (B) function as query panels, offering various querying methods and incorporating heatmaps for enhanced data exploration; (C) is the information panel, presenting pertinent data; (D) features three interactive storyboards for examining the trajectories, patterns, and distributions within the data sourced from the MEI database.}
 \label{fig:DanteXploreVis}
 \vspace{-0.15in}
\end{figure*}

The \emph{DanteExploreVis} has been developed to enable DEs to explain, explore, and present MEI data. The interface is organized into four  panels (Fig.~\ref{fig:DanteXploreVis}). Panels A and B present data through heatmaps, facilitating data exploration, global observation, and data search. Panel D serves as the primary data observation panel, offering map-based visualizations to display book movement and provenance information from various angles. The information panel C presents the search results from Panels A and B, allowing users to highlight and identify specific records within Panel D. The tool is implemented using d3.js~\cite{d3js} and Leaflet~\cite{leaflet_api}.
\subsection{Query Panels with Exploratory Heatmaps}
The Query Panels (A \& B) allow users to perform initial data exploration and make queries. Different types of queries are enabled to obtain the desired MEI records and related visualizations:

\textbf{Instances Heatmap Query Panel} (a1) is based on the flattened transfer data in \textbf{D4}. It focuses on capturing the frequency and distribution of pathway locations and visualizes transfers as individual units rather than as a child layer of a single copy. The heatmap matrix is constructed using the origin-destination of the flattened transfers. Users can select a specific start and end country pair, and click on the corresponding cell to get all the copy records with one or more transfers that meet the query requirement. Two ordering methods are supported: 1) frequency-based ordering enables the user to detect regions where most transfers occurred; 2) alphabetical ordering of country names provides convenience for the user to identify the country and make a selection, serving a better querying role. Hover-on functions are available for each cell, providing a detailed textual description of each cell's data.

\textbf{Full Journey Query Panel} (a2) is designed to search for copy records with known destinations. The copy serves as the basic unit, and the location of the last provenance is the key feature for searching. Given a location pair (i.e., the copy was printed and its current location), the output will be all copy records printed in the same place, traveled for centuries, and eventually ended up in the same destination.

\rev{\textbf{MEI ID Search} (a3) is designed to search specific records with known IDs, with the corresponding visualizations show up immediately.}

\textbf{Time \& Location Heatmaps} (b1 \& b2) serve mostly as exploration tools, providing both chronological and geographical overviews of the dataset. Each row in the heatmap represents an individual copy. A cell in the heatmaps can serve as a query trigger to return the corresponding MEI record. Hover-on functions for detailed information are also supported. The time-focused heatmap (b1) sorts and counts all provenance instances contained in a copy according to the corresponding historical period. Users can easily identify periods with frequent book transfers. The location heatmap takes the spatial feature of each provenance into account. Every cell depicts how many times a book has stayed in a country, allowing users to easily obtain information about the most popular countries with a high number of book transfers and stays.
\subsection{Information Panel} 
The Information Panel (C) acts as connecting component for the entire tool. It links data exploration with data interpretation and bridges the functions of distant and close reading of the data. The information panel displays: 1) in c1, the user-selected query category, the detailed query content, and general statistics of the results, and 2) in c2, a list of the buttons with MEI IDs of the selected copies. The buttons act as the entry point to the three storyboards for close reading and exploration of the data. Clicking a button anchors the visualization of the selected copy in the \textbf{\textit{MSS}} (d1) and highlights its full path in the \textbf{\textit{SSS}} (d2).

\subsection{Storyboards} 
Storyboards (D) are the primary feature supporting data exploration and explanation through multiple functionalities. Three different storyboards have been developed, using a combination of static and dynamic visualization techniques to effectively represent the movement history of copies both temporally and spatially, catering to the needs of DEs.

\textbf{Multi-Static Storyboard} (d1), or \textbf{\textit{MSS}}, emphasizes visualizing individual copy records. Each copy's visualizations are organized in a row, enabling easy navigation and comparison between copies. 
Visualizations for each copy are divided into two groups: an overview and support of close-reading analysis.
Three components are designed to overview the data: 1) \textbf{\textit{radar chart}} (d1.1) displays the data completeness levels [\textbf{D3}], allowing users to quickly gauge the number of provenances contained in a copy and the accuracy of the data (start and end time, and location of stay) for each provenance; 2) \textbf{\textit{full journey map}} (d1.3) visualizes the provenances and transfers of a copy on a geopolitical map using circle markers and arcs with gradient colors to show the progression of time. Zooming in and trajectory animation are supported on the map; 3) \textbf{\textit{horizontal journey}} (d1.2) compresses the geographic location dimension from 2D to 1D and focuses on displaying the progression over time. This element also acts as a connection to \textbf{\textit{step-by-step sequence map}} (d1.4). By clicking on a circle marker, users are directed to the corresponding detailed map. When employing the \textbf{\textit{Instances Heatmap}} (a1) for queries, the transfer inquired will be highlighted in the horizontal journey.
The \textbf{\textit{step-by-step sequence maps}} (d1.4) are designed for the close reading of each transfer, consisting of a cohesive series of geopolitical maps that present the specifics of every transfer. To maintain visual consistency, circle markers and polylines with gradient colors are employed to depict the sequence of provenances and transfers. Each map in the series focuses on the two most recent points, ensuring users can access granular geographic information for every transfer and easily determine if the subsequent provenance stays in the same location. A donut chart glyph, with colors consistent with the radar chart, is integrated into each circle marker to indicate data completeness for the provenance. The Ant Path animation~\cite{react-leaflet-ant-path} highlights the most recent movement path and its direction.
\textbf{Single-Static Storyboard} (d2), or \textbf{\textit{SSS}}, is designed to display trajectories of all copies on one map. To mitigate visual clutter, particularly with large datasets, users can opt for bundled paths using the force-directed bundling algorithm~\cite{holten2009force}. 
\textbf{\textit{Straight path}} employs line segments to intuitively represent locations and routes, enabling users to grasp location and path distribution. In contrast, the \textbf{\textit{bundled path}} conveys similar data but diminishes visual clutter and facilitates the detection of high-frequency routes and locations associated with numerous transfers and stays.
By clicking the buttons in \textbf{\textit{information panel}} (c2), users can highlight trajectories related to specific copies using Ant Path animation, making it easier to identify individual copy trajectories and movement directions of particular paths.
\textbf{Single-Dynamic Storyboard} (d3), or \textbf{\textit{SDS}}, is designed to display the circulation of selected copies through animation on a single map. The visualization animates each copy's movements, illustrating the sequential path and direction for every segment. Two animation modes are available: 
1) \textbf{\textit{all-at-once animation}} begins movement for all selected copies simultaneously upon clicking the play button, 
2) \textbf{\textit{one-by-one animation}} starts each copy's animation sequentially, using distinct colors for differentiation. Users can start, pause, resume, or stop the animation.
A reverse ID lookup function is provided for each trajectory. In the all-at-once mode, a blue pop-up button is linked with each moving marker, allowing users to identify the specific copy of the MEI ID, and clicking on the button, will redirect users to its MEI record page.
%
\section{Evaluation}
\label{sec:Sec8.1}

\begin{figure}[t!]
 \centering 
 \includegraphics[width=0.82\columnwidth]{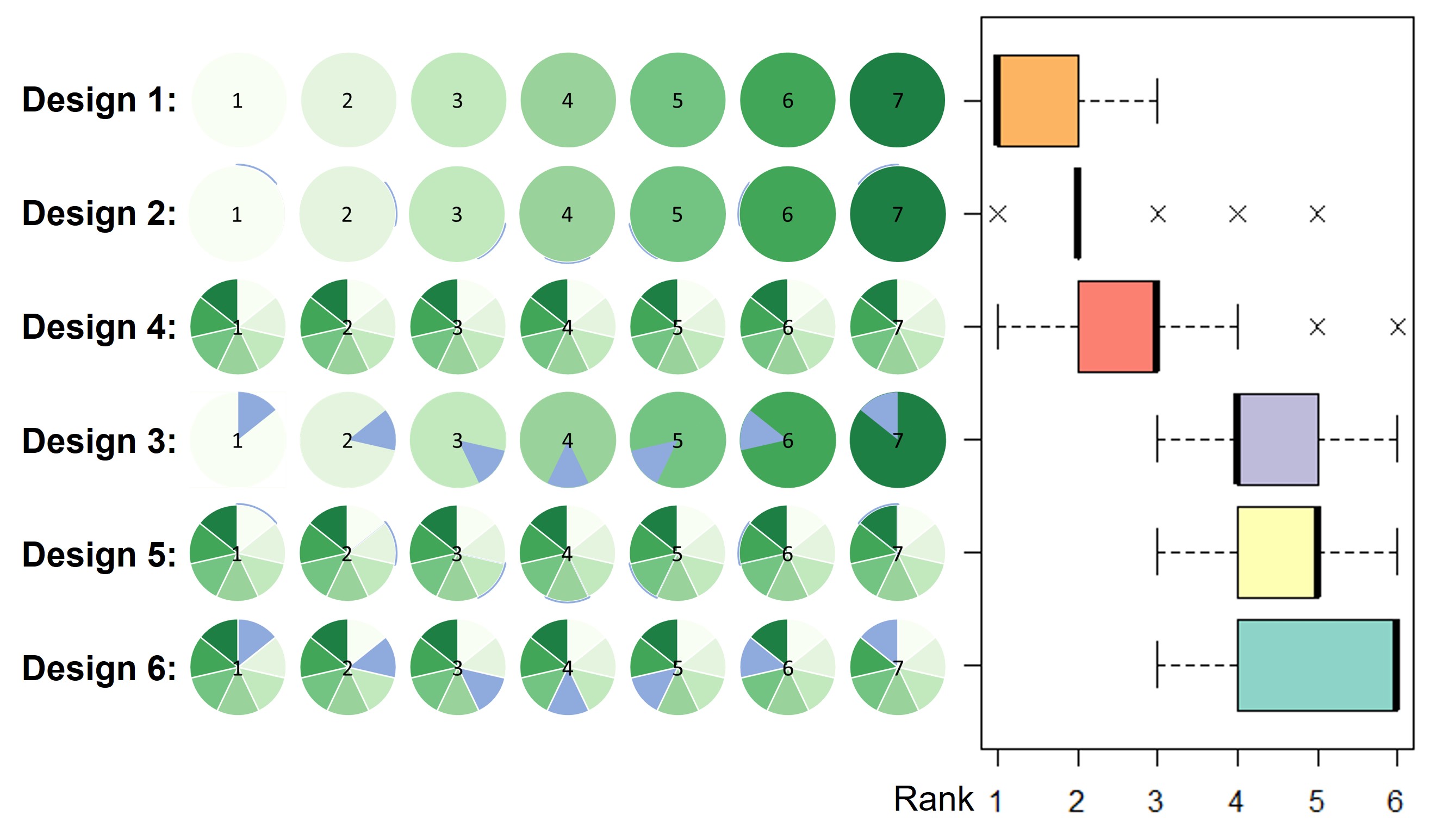}
 \vspace{0.1in}
 \includegraphics[width=0.82\columnwidth]{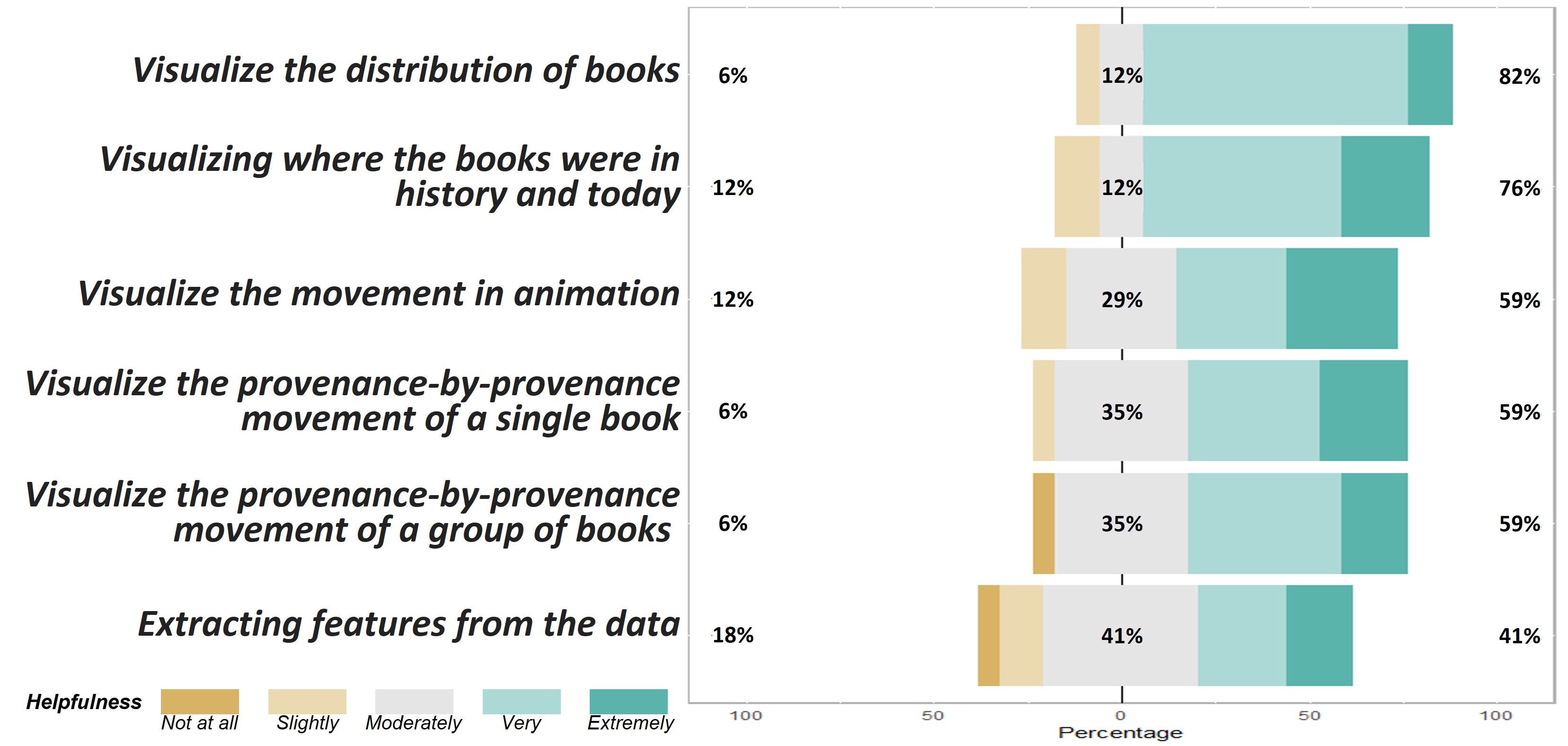}
 \caption{Results of participants' ranking of the six glyph designs (top) and statistics of the Likert scale question responses (bottom). }
 \label{fig: evaluation results}
 \vspace{-0.15in}
\end{figure}

Multiple evaluation methods were employed throughout the design study.  
During iterations, we refined the design via feedback from the LDE. 
The validations in each iteration are detailed in Section~\ref{sec:Sec6}.
In this section, we discuss the evaluation involving a broader group of \rev{DEs} after completing most of the design. The evaluation involved a user survey and expert interviews. The objectives encompassed verifying the representativeness of summarized requirements among \rev{DEs}, investigating the acceptance of various design alternatives, and assessing the effectiveness of the tool in addressing domain requirements.

\subsection{\rev{User survey}}
The user survey aimed to: 1) verify if the majority of domain users concur with the distilled domain requirements obtained through discussions with the LDE, 2) gather feedback on design alternatives for specific tool components, 3) collect quantitative data on the tool's usability and usefulness, and 4) identify potential experts for follow-up studies. Given that \emph{DanteExploreVis} targets historians using the MEI database as their primary research data, we obtained a mailing list of historical book researchers from the LDE and used Expert Sampling~\cite{etikan2016comparison, etikan2017sampling} for participant recruitment. The survey was conducted using Qualtrics~\cite{Qualtrics}.
After the pilot tool went live, we distributed a questionnaire to a \rev{DE} mailing list. The detailed questionnaire is provided in the supplementary material. Out of 45 responses received, 40 were from domain users working in fields related to historical book trading and using MEI data. Among these participants, 33\% had used visualization tools in their work, 93\% believed that visualization tools could support their research and expressed interest. In summary, our survey findings are as follows:

\textbf{\textit{Domain requirements are representative.}}
We involved the LDE throughout the design study, who is the creator of MEI and has experience in medieval book research and teaching. However, consulting a wider range of DEs is crucial to ensure rigor and verify that the summarized requirements accurately represent the domain's needs.
The survey has provided confirmatory results: 90\% of the participants think it would be helpful to visualize the full circulation of a single book \& a group of books on a geopolitical map (\textbf{R1, R2, R3}). 85\% of them would like to see an animation showing the transfer of books on the map (\textbf{R5}). 83\% show the desire to see the features extracted from the transfer instances in a single visualization (\textbf{R6}).

\textbf{\textit{Keep designs simple and connected.}}
During the design process, most design decisions were made in agreement with the LDE, but some remained unresolved. These primarily focused on visualizing the chronological order of circles on the Full Journey map in the \textbf{\textit{MSS}}. We suggested two gradient-based circular filling options to represent progress: 1) a single color from the gradient color group, and 2) the entire gradient color range for conical filling. For each method, we provide three options to highlight the current stage. Survey feedback (Fig.~\ref{fig: evaluation results} \textit{top}) aligned with the LDE's opinion. Single-color filling ranked first. The feedback indicated that simplicity and comprehensibility were more important than displaying more information.
Conical filling for small circles sometimes distracts from the overall map view. Connections to other visualizations
, can compensate for the concealed information when using single-color filled circles.

\textbf{\textit{The tool is useful.}}
Concerning domain requirements, we examined the tool's usefulness. The 5-point Likert scale ratings indicated positive outcomes (Fig.~\ref{fig: evaluation results} \textit{bottom}). 

%

%
\subsection{Expert Interview}
The features in \emph{DanteExploreVis} were further refined and improved based on survey feedback and then deployed. By conducting expert interviews along with the think-aloud protocol, the tools' effectiveness and usability were evaluated. We observed DEs' interaction with the tool and performed qualitative analysis using transcripts.

\subsubsection{Methodology}
We interviewed 4 DEs recruited from the survey. Each interview lasted between 60 $\sim$ 90 mins.
%
All participants use MEI for research \rev{for more than two years. P2 (2 years) is a PhD student involved in the Dante Project copy census; P3 (6 years) is an initial builder of the MEI database, contributing to record expansion and verification. P1 (5 years) and P4 (2 years) also use MEI for teaching.} 
%
%
\rev{The interview combines think-aloud evaluation and reflective discussion for multifaceted user-interaction insights.} 
%
It started with a demo then participants screen sharing and thinking aloud during the prepared activities. They first explored the tool, then completed tasks based on pre-collected domain requirements.
The interview ended with discussions on tool usability, effectiveness, and potential development. 

\subsubsection{Results}
%
%
\textbf{Reflection on Usability:}
Drawing on Nielsen's five usability indicators\cite{nielsen1994usability}, feedback from DEs indicated good levels of \textit{satisfaction} and \textit{efficiency}, while \textit{memorability} was not the primary evaluation focus. Concerning \textit{learnability} and \textit{low-error}, most participants quickly understood and correctly used various features, though some initially struggled with heatmaps.
P1 and P4 had questions about cell meanings in the \textbf{\textit{Instances Heatmap}} (Fig.~\ref{fig:DanteXploreVis} a1). They 
found it challenging to view provenances and transfers as separate entities, regardless of the associated copy. After guidance, they recognized its value in providing an overview and highlighting unique traits. P1, P3, and P4 commented on the small font size of axis labels. To enhance heatmap usability, we added hover-over with more 
descriptions and increased the font size.


%
\textbf{Reflection on Usefulness:}
Usefulness was evaluated based on the tool's contribution to meeting domain requirements. 
All participants found the map-based visualization helpful in displaying the geographical distribution of books (\textbf{R3}). According to transcripts, various features in the tool facilitated intuitive data interpretation. New data manipulations and corresponding visualizations allowed efficient exploration (\textbf{R1, R2, R4}) and provided insights into the prevalent book transfer causes across time and space (\textbf{R6}).
%

%
The \textbf{\textit{Storyboards}} received wide appreciation for their ability to present and explain the transfer history of book copies on map-based views from multiple perspectives. Participants P1 and P3 mentioned that \textbf{\textit{MSS}} (Fig.~\ref{fig:DanteXploreVis} d1) presented individual copies as row objects aligned with their accustomed way of viewing the records in the MEI interface. The linked global and partial views were very helpful in quickly understanding the copy's movement history and exploring provenances with interesting features. All participants commended \textbf{\textit{SSS}} (d2) for effectively displaying the distribution and aggregation of transfers across all printed copies of one book edition, providing previously inaccessible information. As an animated version of the \textbf{\textit{SSS}}, the \textbf{\textit{SDS}} (d3) was praised for the vivid presentation of book movements.
The \textbf{\textit{Heatmap}} was acknowledged as an efficient method for identifying and searching interesting patterns in MEI. Participants noted their intuitive presentation of transfer data characteristics across all copies regarding geographical or temporal distribution. Linking \textbf{\textit{heatmaps}} to \textbf{\textit{storyboards}} enabled quick targeting of specific copy records with certain traits for closer observation: \say{\textit{It is a tool that cuts the time of realizing things and will save you a lot of comparing and looking at data while you have it all here.}}
Participants P2 and P3 found heatmaps especially useful for providing a new perspective for users with limited prior knowledge, offering insights and research starting points: \say{\textit{The heatmaps are useful at the start of a research when someone wants to immediately see a pattern or behavior, or what they need to focus on.}}
The concept of \textbf{\textit{data completeness}}, obtained through continuous data abstraction during the iterative design process, was well-received by participants. They appreciated the \textbf{\textit{radar chart}} and \textbf{\textit{step-by-step sequences map}} (Fig.~\ref{fig:DanteXploreVis} d1.1 \& d1.4) for providing a quick overview of data ambiguities in each provenance and an indication of credibility and reliability. Participants also liked the 
connection to MEI page for data checking and proofreading. The animations embedded in the tool received positive feedback, particularly from DEs with teaching duties, who noted their usefulness in engaging students and showing patterns.

%

%
\textbf{Usage Scenarios:} We described the following scenarios encountered or envisioned by DEs during the think-aloud evaluation and reflective discussion. 
P2 was familiar with all the copy records entered for the Dante Project and wanted to use the tool to provide evidence for her research conjectures. She started with the temporal heatmap (Fig.~\ref{fig:DanteXploreVis} b1) and observed the frequency distribution of transfer activity in the temporal dimension for all copies: \say{\textit{It's very interesting to see that the colors are darker in these beginning time frames and then get lighter and less frequent. It illustrates how the Printing Revolution in Early Modern Europe impacted the book trade, [...], I can visually see they increase again around the 18${^{th}}$ century, which can be used as evidence of the mechanization of printing technology.}} With a particular interest in the copies that had traveled across the ocean to the US, she opened the Full Journey query panel (a2), set the start and end location pair, and got 25 records retrieved. She then opened the \textbf{\textit{SSS}} (d2) for global exploration. She identified and compared the transfer paths of different copies on the map by clicking on the corresponding buttons in the Information Panel (c2). In the end, from the \textbf{\textit{SDS}} (d3), she enjoyed the animation of the selected copies moving across the map: \say{\textit{It is surprising to see these books printed from the same place and headed to the same destination but underwent such different journeys.}} 

P4, focusing on the special immigration of books from Europe to South America, was less familiar with the MEI records entered for the Dante Project. He began exploring using the Instances Heatmap (a1), ranking results by frequency, and noticing most transfers were in Europe. Focusing on the UK, he obtained all the copies that contained transfers within the UK by clicking the cell. Using the \textbf{\textit{MSS}}, he compared UK transfers in different copies with the Horizontal Journey (d1.2). He targeted interesting records by combining Full Journey and Step-by-step Sequences views (d1) and then went for more information on the linked MEI record page. He envisioned applying the tool for his research: \say{\textit{The heatmaps can accelerate the process of seeking the migration of books from Europe to South America, [...], the trajectory and the animation can further show the migration took place in a few decades and how the new migration tool place.}} 
%

%
\textbf{Summary}
Fulfillment of \emph{DanteExploreVis} requirements was confirmed during interview sessions. Heatmaps' usability issues were addressed based on participants' feedback.
On usefulness, all domain requirements were met. The three storyboards provided a multi-perspective interpretation of MEI data in static or dynamic forms, fulfilling requirements \textbf{R1}, \textbf{R2}, and \textbf{R5}. Map-based visualizations addressed the research gap (\textbf{R3}), complementing textual MEI records and supporting conjectures. One participant remarked, \say{\textit{The tight link to the map is really useful. This tool proves a saying that geography is the mother of history.}} The data completeness proposal and associated visualizations improved data checking efficiency (\textbf{R4}). Heatmaps and the hierarchical provenance data's disintegration and integration offered new exploration avenues, inspiring further research (\textbf{R6}).

%

\section{Reflection and Discussion}
\label{sec:Sec8.2}

\rev{Reflecting on our experiences, we highlight the unique challenges and opportunities of collaborating with humanities domain experts. We further explore how our adapted framework assists in mitigating these challenges and discuss its potential applicability across other domains.} 

\textbf{\textit{Tailoring design study frameworks for different domains:}}
Our experience underscores the importance of adapting design study frameworks according to the target domain. Experts from different fields possess distinct habits, including information processing, data perception, thinking styles, collaboration methods, etc. Adapting to these habits and customizing the framework for design research implementation can enhance the process's efficiency. For example, historians may possess a deeper understanding of data and tend to present new requirements each time they encounter a new way of visualizing the domain data. As a result, we adjusted the core phase in the Nine-Stage Framework to an iterative design cycle with fixed stages, emphasizing data, requirement, and task reflection during each iteration. 

\textbf{\textit{Balancing engineering and creative design processes~\cite{mckenna2014design}}}: We discovered that humanities researchers exhibit different research habits and sensitivities compared to their scientific counterparts. For instance, they favor textual information and may display lower sensitivity to visual and numerical data. Although they may establish goals and problems to solve at a project's outset, their strong divergent, and creative minds can generate needs unrelated to the current design mid-project. Thus, during the design research process, we aim to fulfill domain requirements; however, if they are too unrelated, we evaluate whether they should be implemented in the current tool or serve as a starting point for future projects or tools.

\textbf{\textit{Being aware of the ``wow'' effect}}: Keep in mind that the wow effect, such as the use of animation or the simultaneous display of vast amounts of data, may sway domain experts during the design process. This could potentially divert their focus from assessing the effectiveness of the tool in supporting data analysis. When examining data from qualitative evaluations, it is essential to prioritize feedback regarding the tool's functionality and overall usefulness. Be discerning about whether praise from domain experts may be affected by the wow effect.

\textbf{\textit{Questioning domain expert assumptions:}} Visualization experts should avoid assuming that domain experts always possess a deeper knowledge of data characteristics and uses. By challenging domain expert opinions and investigating alternative visualization techniques, new research opportunities, and more effective design solutions can emerge. This process can also provide novel perspectives for domain experts to understand or work with the data. For instance, in our case, despite initial skepticism from domain experts about the usefulness of a heatmap, as the project evolved, they discovered its value in interpreting large datasets and as a source of inspiration for their research.

\rev{\textbf{\textit{Overcoming Interdisciplinary Collaboration Challenges:}}}
\rev{The framework variant emphasizes iterative design cycles and deep collaboration with domain experts, cultivating a more context-sensitive and user-centered design perspective. This methodology helps to navigate and reconcile the understanding gaps and thought divergences often found in collaborations spanning distinct domains. Hence, the framework proves to be a versatile instrument for conducting design studies across a broad spectrum of fields, extending its value beyond the humanities domain, which was the genesis of our project.}

\section{\rev{Limitations and Future Work}}
\label{sec:Sec8.3}


\rev{\textbf{\textit{Data uncertainty and ambiguity}}, inherent characteristics in humanities and historical datasets, pose challenges to accuracy. The MEI data is heavily reliant on expert annotations based on physical evidence, thus improving precision via computational means is difficult. Nonetheless, future work will focus on visualizing these uncertainties, attracting expert attention, and potentially improving data quality.}

\rev{\textbf{\textit{Scalability}} is a limitation due to \emph{DanteExploreVis}'s specialized design and integration with the MEI database and data structure. However, the hierarchical and temporal-spatial nature of the provenance data shares similarities with datasets in various fields. Multi-views with close and distant observations applied in \emph{DanteExploreVis} meet the requirements of a visual analytics tool. With minor adjustments, it could be employed for research involving diverse datasets such as population migration, theological book circulation, and the trade of manuscripts and paintings. Testing \emph{DanteExploreVis} with different datasets will be part of our future endeavors to gauge its scalability.}

\rev{\textbf{\textit{Scalable Data Visualization}} is a significant issue. The tool currently accommodates the existing data effectively, the anticipated expansion of the MEI database and the escalating complexity of analytical demands could potentially result in visual clutter. Although our tool includes edge-bundling to mitigate clutter, its future efficiency is uncertain. Future work will explore enhancing the bundling feature by enabling user customization for more targeted path bundling. We are also examining alternative visualization methodologies for temporal-spatial path data to overcome the constraints of a 2D map. These advancements would require continued collaborative discussions and alignment with DEs.}

\section{Conclusion}
\label{sec:Sec9}

We modified the core phase of the Nine-Stage Framework to develop a tailored design study framework for historians. Over six-month, we collaborated with domain experts in iterative design cycles to create \emph{DanteExploreVis}. We thoroughly documented the process to ensure reproducibility and shared insights and lessons learned from our interdisciplinary approach, highlighting its potential for broader application across various domains.
On scalability, the \emph{DanteExploreVis} interface can be adapted to support trajectory data analysis in different fields, though future evaluations with diverse datasets are necessary. At present the tool is applied to a subset of the MEI data; future research could explore its application to the complete dataset or examine relationships between different versions or literary works based on provenances.

\acknowledgments{
Y. Xing is funded by: King's-CSC PhD Scholarship programme.}
\bibliographystyle{abbrv-doi-narrow}

\bibliography{ms}

\end{document}